\documentstyle[12pt,preprint,aps,floats,epsf]{revtex}

\tighten

\begin{document}

\newcommand{\newc}{\newcommand}

\newc{\gsim}{\lower.7ex\hbox{$\;\stackrel{\textstyle>}{\sim}\;$}}
\newc{\lsim}{\lower.7ex\hbox{$\;\stackrel{\textstyle<}{\sim}\;$}}

\def\beq{\begin{equation}}
\def\eq{\end{equation}}
\def\eeq{\end{equation}}

\preprint{
\noindent
\hfill
\begin{minipage}[t]{3in}
\begin{flushright}
UPR-844-T \\
RU--99--17 \\
hep-ph/9905252  \\
May 1999 \\
\vspace*{1.5cm}
\end{flushright}
\end{minipage}
}


\title {A low-energy solution to the $\mu$-problem in gauge mediation}
\author{Paul Langacker$^{\,a}$, 
Nir Polonsky$^{\,b}$,
and Jing Wang$^{\,a}$}
\address{$^{a}$
 Department of Physics and Astronomy, The University of Pennsylvania\\
Philadelphia, PA 19104-6396, USA}
\address{$^{b}$
 Department of Physics and Astronomy, Rutgers University\\
 Piscataway, NJ 08854--8019, USA}
\maketitle

\begin{abstract}
In the gauge-mediation framework the soft supersymmetry breaking
mass parameters of the supersymmetric standard model are induced by the
gauge interactions of some messenger fields.
The parameters exhibit flavor universality
which is dictated by the gauge interactions and which efficiently eliminates
new dangerous contributions to flavor changing neutral
currents. However, the Higgs potential in this framework typically contains
an unacceptable hierarchy between its dimensionful parameters
(the $\mu$-problem of gauge mediation). We show
that the problem can be resolved if the Higgs potential arises
dynamically once an intermediate $U(1)$$^{\prime}$ sector is integrated out
rather than arising radiatively from some Yukawa interactions at the messenger scale.
As an added benefit, such models may naturally avoid new contribution
to CP violating amplitudes.
The proposed framework is described, explicit examples 
are given and its phenomenology is explored. 
The $\mu$ problem is resolved  in this case by the
low-energy $U(1)^{\prime}$ 
dynamics which could be tested in future collider experiments.
\end{abstract}

\newpage
\setlength{\parskip}{1.2ex}

\section{Introduction}
\label{sec:intro}

Supersymmetry provides a well motivated framework for embedding
and extending the Standard Model of strong and electroweak interactions
(SM): Its boson -- fermion symmetry resolves the hierarchy problem
and allows one to consistently extrapolate the theory to high-energy.
However, a long standing question remains:  
At what high-energy scale does the next level of structure reside.
The answer depends in part on the manner in which supersymmetry
breaking is mediated (at a scale $\Lambda_{\rm mediation}$)
from some (hidden) sector in which supersymmetry
is broken spontaneously to the SM (observable) sector. It also
depends on whether the SM sector itself contains any additional
structure, e.g., SM $\times$ $U(1)$$^{\prime}$.

It is possible that the mediation of supersymmetry breaking is carried
out through the SM gauge interactions (in which case there may not be a truly
hidden sector in the sense that it interacts only gravitationally with
the observable sector). Finite gauge quantum corrections involving 
supermultiplets of postulated heavy (vector-like) matter which
transform under the SM  -- {\it the messengers} -- generate
the desired soft supersymmetry breaking
(SSB) mass parameters in the low energy potential. 
One has in this case $m_{\rm SSB} \sim 
(\alpha/4\pi)\Lambda$, where $\Lambda$ is a scale\footnote{
The messenger scale may be
roughly the same as or smaller by a few orders of magnitude than the actual
scale of supersymmetry breaking.}
of the order of magnitude of the masses of the messengers
(it is given by the ratio of a supersymmetry breaking 
mass-squared and a supersymmetry conserving mass) 
and $\alpha$ is a generic gauge coupling. 
This is the gauge-mediation framework \cite{DNS}.
The physical high-energy scale in this case is set by
$\Lambda \sim (4\pi/\alpha)m_{SSB} \sim (4\pi/\alpha)(1 - 10) \times M_{W} 
\sim 10^{4 - 5}$ GeV, where $m_{SSB}$ is a typical (SM) superpartner mass.
A clear benefit of this framework is the flavor
universality of the sfermion $\tilde{f}$ spectrum, which comes about since the
different $m_{\tilde{f}}^{2} \sim Q_{f}^{2}[(\alpha/4\pi)\Lambda]^{2}$ 
parameters are distinguished only by the respective 
gauge quantum numbers. This eliminates {\it a priori} many potentially
dangerous sources of flavor changing neutral currents.

The low-mediation scale $\Lambda_{\rm mediation} \simeq \Lambda \sim 10^{4-5}$ GeV
renders negligible, in most cases, any supergravity
effects and hence eliminates various supergravity schemes for the 
generation of the $\mu$-parameter -- 
the supersymmetry conserving Dirac mass which mixes
the two Higgs doublets of the supersymmetric extension (SSM)
$W = \mu H_{1} H_{2} + \cdots$. Hence, the $\mu$ problem,
why $\mu \simeq {\cal{O}}(M_{W})$ rather than $\mu \simeq {\cal{O}}
(\Lambda_{\rm mediation})$, resurfaces in this case. 
(The non-observation of chargino pairs at the
$WW$ threshold implies that $|\mu| \gtrsim M_{W}$, and in
particular, that it cannot vanish \cite{FPT}.) One can consider the
possibility that the dimension-one $\mu$-parameter is generated by 
Yukawa quantum corrections which involve some messenger  fields
which interact with the Higgs
doublets of the SSM via Yukawa couplings $y$. However, if this were the
case then it is straightforward to show \cite{DGP} that a dimension-two SSB
mixing between $H_{1}$ and $H_{2}$ in the scalar potential $V_{SSB}
 \sim \cdots + m_{3}^{2}H_{1} H_{2} + h.c. + \cdots$ would also be
generated and at the same loop order. 
Unlike in the case of gauge loops, dimension one and two parameters
generated by Yukawa loops are typically suppressed by the same power of the loop factor.
As a result $|\mu| \sim (y^{2}/16\pi^{2})\Lambda \sim m_{3}^{2}/\Lambda \sim M_{W}$
(or equivalently, $\mu^{2} \sim (16\pi^{2}/y^{2})m_{3}^{2}$).  
It reintroduces a hierarchy
problem to the Higgs potential which would be dominated in this case by
$m_{3}^{2} \sim M_{W}\Lambda$. This is a most severe problem
that undermines any success of the gauge-mediation framework.

The most successful attempts to address this new hierarchy problem
involve in one fashion or another the details of 
the high-energy (supergravity) theory, and in that sense
they are high-energy solutions. For example, Ref.~\cite{DGP} invokes 
a radiative linear term generated by messenger-scale singlet interactions.
The linear term shifts a singlet field $N$ (which interacts with the
Higgs doublets) to a scale which is
suppressed by a loop factor in comparison to the messenger scale.
The shifts in the scalar and auxiliary
components of $N$, which induce $\mu$ and $m_{3}^{2}$ respectively,
arise at different loop orders, evading the above described hierarchy
problem.  The superpotential (or equivalently -- the K\" ahler potential) 
couplings must be fixed by the  high-energy ($Q > \Lambda$) theory. In
particular, a scale associated with a tree-level 
linear term must be fixed to be ${\cal{O}}(\Lambda)$.
Alternatively, in Ref.~\cite{NP} it was pointed out that a radiative linear term
in a singlet field $N$ is typically generated by supergravity 
and is suppressed by only one inverse power of the Planck mass $M_{P}$.
Hence, it can still play an important role in the low-energy theory.
It shifts the singlet field $N \sim (\Lambda^{4}/\kappa^{2}M_{P})^{1/3}$
(assuming $W \sim (\kappa/3)N^{3}$).
The singlet Yukawa interaction with the Higgs doublets then generates
the desired parameters at tree-level $\mu^{2} \sim m_{3}^{2} \sim N^{2} \sim
M_{W}^{2}$ (assuming that supersymmetry is broken at a scale of the
order of $10^{6\pm1}$ GeV). 
In this case no new scales are introduced by hand, but there is still
dependence on the high-energy theory. 
(A somewhat similar application of
supergravity to the problem was proposed in Ref.~\cite{Y}.)
Both solutions assume the presence of a ``dedicated''
messenger-scale gauge singlet(s), denoted above by $N$.

Here, we point out a distinctive possibility that the singlet field is not a
gauge singlet but only a SM singlet $S$. Specifically, we assume
the extension $(S)SM$ $\rightarrow$ $(S)SM$ $\times$ $U(1)$$^{\prime}$, and  
that $S$ carries a charge $Q_{S} = - (Q_{H_{1}} + Q_{H_{2}})$ under
the  additional Abelian symmetry so that a Yukawa 
term $W\sim h_s S H_{1}  H_{2}$ is allowed. In turn, a scale 
$\Lambda^{\prime} \sim \langle S \rangle \lesssim \Lambda$, which is
associated with the breaking of the $U(1)$$^{\prime}$, must be
introduced, or preferably, induced.
The various $\mu$ problems of gauge mediation 
will be shown to be solved in this case by the low-energy dynamics
associated with this new scale.

The scale $\Lambda^{\prime}$ could be  generated radiatively and is a function 
in this case of 
$\Lambda$ and of ${\cal{O}}(1)$ Yukawa couplings. A coupling between $S$
and exotic quarks, e.g., $D$ and $D^{c}$ singlets with hypercharge
$\pm(1/3)$, generates negative
corrections to $m_{S}^{2}$ so that $m_{S}^{2}(\Lambda^{\prime}) < 0$ and 
$S$ acquires a vacuum expectation
value (vev). This is essentially a $U(1)$$^{\prime}$ version of the
well-known radiative symmetry breaking (RSB) mechanism that 
is responsible in the SSM for the generation of 
the negative mass term in the SM Higgs potential. 
A similar  idea was 
suggested previously in the context of supergravity and high-energy
(gravity) mediation ($\Lambda_{\rm mediation} \simeq M_{P}$)
of supersymmetry breaking \cite{CL}.
In that case, like RSB in those models, 
the large evolution interval enables one to render 
$m_{S}^{2} < 0$ somewhere above the weak scale. 
In the supergravity case the superpotential interactions generate
$|\mu| \sim h_s \langle S \rangle$
while trilinear SSB terms
$V_{SSB} \sim \cdots + h_s A_s S H_{1}  H_{2} + h.c. + \cdots$ generate
$m_{3}^{2} = A_s h_s \langle S \rangle $. Since all parameters 
in the gravity-mediation framework
are of the same order of magnitude as the gravitino mass (which is
fixed $m_{3/2}\sim M_{W}$),
then $h_s \langle S \rangle $ is expected to be of the same order of magnitude as well.
This leads to  a successful solution to the $\mu$-problem
in high-energy supergravity models. 
Various other solutions that benefit from the large mediation scale
are also available in the supergravity framework \cite{sugra}.
The radiatively broken $U(1)$$^{\prime}$ scenario was extensively
studied in the framework of supergravity and of string-inspired models 
\cite{penn}.

In contrast to the supergravity framework, 
in gauge mediation the evolution interval is short; in addition,
trilinear parameters are highly suppressed\footnote{Large $A$
parameters were proposed, however, in Ref.~\cite{P}.} 
$A \sim (\alpha/4\pi)^{2}\Lambda\ln\Lambda$.
While the small $A$ parameters remain a constraint,
the shorter evolution interval
is more than compensated (as for the case of RSB in these models) by the large
hierarchy within the SSB parameters
$m_{D}^{2}/m_{H}^{2}/m_{S}^{2} \sim
\alpha_{3}^{2}/\alpha_{2}^{2}/\alpha^{2}_{1^{\prime}}$
(where $\alpha_{3,\,2,\,1,\,1^{\prime}}$ are the
$SU(3),\,SU(2),\,U(1)$, and $U(1)^{\prime}$ gauge couplings). In fact, the
messengers may not transform under $U(1)^{\prime}$, in which case 
$m_{S}^{2}(\Lambda) = 0$. For $\alpha_{1^{\prime}} =
{\cal{O}}(\alpha_{Y})$, which we will assume,
the exact boundary condition for $m_{S}^{2}$ does not affect our discussion
and for simplicity we assume hereafter that the messengers are indeed
invariant under $U(1)^{\prime}$.

A radiatively induced  $\langle S \rangle$ as a source of $\mu$
in the case of a gauge singlet $S$ was considered previously
in the context of gauge mediation~\cite{singlet}.
It was found that the singlet must couple to exotic quarks
with large Yukawa couplings, as naturally occurs in the context of
$U(1)$$^{\prime}$. In the gauge singlet case, however, the
superpotential must contain a $S^{3}$ term so that the potential contains
quartic terms $V \sim |\partial (S^{3} + SH_{1}H_{2})/\partial S|^{2}$
which stabilize it.
The models suffer from
the usual problem of a spontaneously broken global $Z_{3}$ symmetry 
(under which $S^{3}$ is invariant) which
results in unacceptable domain walls at a (post-inflationary) low-energy epoch.
In the gauged case $S$ is not a singlet and  
$S^{3}$ terms are not gauge invariant and are automatically
forbidden.  Instead, the potential is stabilized by $U(1)$$^{\prime}$ 
gauge $D$-terms $V \sim  \cdots + (g_{1^{\prime}}^{2}/2)(Q_{S}|S|^{2} +
Q_{H_{1}}|H_{1}|^{2}  + Q_{H_{2}}|H_{2}|^{2})^{2} + \cdots$
(which are not available for a gauge singlet $S$).
The $Z_{3}$ symmetry is now only a (harmless) subgroup of the gauged 
$U(1)^{\prime}$. 
While in the non-gauged case
the former source of the quartic terms also generates an additional
contribution to $m_{3}^{2} \sim
S^{2}$, this is not possible in the gauged case 
(with only one singlet).

In either the gauged or non-gauged case, 
the potential also exhibits an approximate phase ($R$) symmetry, which
exists in models with only Yukawa superpotential terms and corresponds to a
rotation of all fields by the same phase. 
It is broken spontaneously by $\langle S \rangle$ and  explicitly by
tri-linear $A$-terms. The explicit breaking is, however,  suppressed by the
smallness of the $A$-parameters. We will show below that 
in spite of the suppressed $A$-parameters 
it is possible to generate $m_{3}^{2}$ 
and break the phase symmetry strongly enough
to avoid the light pseudo Goldstone boson which otherwise appears.
Specifically, as will be shown below, it is very likely that
in the $U(1)$$^{\prime}$ scenario $M_{W} \ll \langle S\rangle \lesssim
\Lambda$ and hence, $m_{3}^{2} \sim h_s A_s \langle S\rangle
\sim A_{s}\mu$ is a geometric mean of a small parameter and
a large vev. It implies a somewhat  large 
value of $|\mu| \sim  {\cal{O}}(1\, {\rm TeV})$. However, this  
typically occurs in gauge mediation 
as a result of RSB constraints
in the presence of a heavy gluino~\cite{GMRSB}. 
The details depend strongly on the exotic quark spectrum
and on the $U(1)$$^{\prime}$ charges, and some examples will be
presented below. Alternatively, in models with two singlets a superpotential
term $SS^{\prime 2}$ could be gauge invariant, and
$\langle S^{\prime}\rangle \sim \langle S\rangle $
could generate an additional contribution to $m_{3}^{2} \sim \langle 
S^{\prime}\rangle^{2}$, just as in the non-gauged case. 
(Note that in the non-gauged case 
the $U(1)^{\prime}$ rotations -- explicitly broken by the $S^{3}$
terms -- correspond to global transformations 
and there is one additional pseudo Goldstone boson.)

It is particularly interesting to note that in the models
with only one SM singlet there appear only two new phases which can be
rotated away, and hence there are no new physical phases\footnote{
We thank J.~Feng and T.~Moroi for bringing the CP problem
in gauge mediation and its possible solution to our attention.}. 
This is because there is only one common phase to all gaugino mass
and the radiatively-induced $A$ parameters, 
while the phase of $m_{3}^{2}$ is given in this case 
by the phases of $\mu$ and $A$.
Hence, after $R$ and Peccei-Quinn rotations no physical phases
appear in the soft parameters.
This eliminates new
contributions to CP violating amplitudes such as the electron dipole
moment, which are flavor conserving and which generically appear at 
unacceptable levels even in gauge-mediation models \cite{Moroi}.

The stabilization due to the $D$-terms and the generation of the $A$
terms then open the door to new (low-energy) solutions to the
$\mu$-problem in gauge mediation. The mechanism is quite different from 
that of the non-gauged case since the quartic coupling is given, in
principle, by a fixed gauge coupling rather than by a free superpotential
coupling; $m_{3}$ must depend on overcoming the suppression of the
tri-linear couplings $A$; and the scale $\langle S \rangle$ 
is a physical scale with observable consequences.
Hence, it corresponds to a distinctive and interesting option
and it will be explored in some detail below.

The $U(1)$$^{\prime}$ models predict, in addition to the extra matter
and the associated rich spectrum,
an extra gauge boson, $Z^{\prime}$. The corresponding phenomenology
is similar to that of any other model with $Z^{\prime}$, except that
$M_{Z^{\prime}}^{2} \sim -(Q_{S}/2)m_{S}^{2}$ is large,
given that $|m_{S}^{2}|$ is controlled by the large exotic quark
SSB parameters. Typically we find $M_{Z^{\prime}} \simeq {\cal{O}}(1
\, {\rm TeV})$ and with suppressed mixing with the ordinary $Z$-boson.
Thus it decouples safely from electroweak physics \cite{EL}.
Another interesting aspect of supersymmetric $U(1)$$^{\prime}$ models
that repeats here is that the tree-level light Higgs $h_{1}$ 
mass exceeds its usual upper bound of $M_{Z}$.
This is due to contributions from the $U(1)$$^{\prime}$ 
$D$-terms to the quartic potential, which lift its otherwise flat 
direction. We find for its mass $m_{h_{1}} \simeq 120 - 150$ GeV 
at tree level and $m_{h_{1}} \simeq 150 - 180$ GeV at one loop.

A most interesting aspect of the $U(1)$$^{\prime}$ scenario is that the gauge-mediation scale is still the only
fundamental scale, and the $U(1)$$^{\prime}$ scale is determined
from it. It has been proposed recently \cite{CDM} that perhaps the same
$U(1)$$^{\prime}$ is also responsible for the actual mediation
of supersymmetry breaking from the ``hidden'' sector to the messenger
fields (i.e.,  to identify $U(1)$$^{\prime}$ with $U(1)$$_{\rm messenger}$
of Ref.~\cite{DNS}). This is an ambitious yet interesting proposal  that
significantly differs from our bottom-up  approach,  
which, in principle, is independent of the details of supersymmetry
breaking and its initial mediation to the messenger fields.
By distinguishing the two extended interactions we avoid the need, 
e.g., to fine tune
Yukawa couplings, which is the situation in Ref.~\cite{CDM} due to the
multitude of tasks imposed there on a single  $U(1)$. 
The only (moderate) hierarchy in Yukawa couplings that is
assumed is between those that involve (exotic) quarks, which are
taken to saturate or be near their infra-red quasi-fixed points 
and be ${\cal{O}}(1)$, and those
which involve only the Higgs doublets and the singlet(s), which do
not reach any (quasi-)fixed points and hence are taken to be smaller.
Such differences naturally stem from QCD renormalization, which enables
the existence of quasi-fixed points for the (exotic) quark couplings.

We describe the model in greater detail in the next section.
In Section III we discuss the evolution of its parameters, the scalar
potential, and the extraction of $\mu$ and $m_{3}^{2}$
in the two schemes. We summarize and comment on various issues such
as unification in the last section.
Since our focus in this work is on the demonstration of the viability
of the proposed low-energy solution to the $\mu$-problem in gauge mediation,
some assumptions (which are clearly indicated) 
were made in order to simplify the
presentation and discussion (mainly by fixing parameters whose variation was
found to be immaterial for our purposes). Also, only results for the Higgs and
neutralino/chargino sectors which significantly differ from the usual case
will be presented in detail, though we
checked that the complete spectrum is consistent. Our presentation is
within a one-scale model though generalizations are straightforward.
Our focus is on the low-energy theory below the messenger scale and on
the generation of the $U(1)$$^{\prime}$ and electroweak scales as
functions of this scale. We comment on implications for possible
extrapolations to higher energies in our conclusions.

\section{The Model}

The extended gauge symmetry of the model is $ SU(3)_{c} \times 
SU(2)_{L} \times U(1)_{Y} \times U(1)^{\prime}$, 
with the gauge couplings $g_{3}$, $g_{2}$,
$g_{1}=\sqrt{\frac{5}{3}}g_{Y}$ and $g_{1^{\prime}}$ respectively, and
$\alpha_{i} = g_{i}^{2}/4\pi$.
The extended particle content is given
by the left-handed  chiral multiplets of the 
the minimal supersymmetric extension (MSSM)
including three families of quark doublets ($Q$) and singlets ($u^{c}$ and $d^{c}$),
lepton doublets ($L$) and singlets ($e^{c}$), and the Higgs doublets $H_{1}$
and $H_{2}$. In addition, it contains exotic 
quark vector-like pairs, e.g. $D$ and $D^{c}$ pairs 
which are singlets under $SU(2)$$_{L}$ and carry hypercharge $\pm 1/3$, and
fields which are singlets under the SM but are 
charged under $U(1)$$^{\prime}$. We
assume that all of the low-energy matter fields (but not the
messenger-scale fields) are charged under $U(1)$$^{\prime}$. 
For concreteness, we further assume in most examples that the 
$U(1)$$^{\prime}$ charges of the fields are given by the 
$U(1)_{\eta}$ of the $E_{6}$ model \cite{LW}, up to an overall normalization.
The $E_{6}$ symmetry is used as a classification tool only, i.e.,
we do not assume a full grand unification.
Nevertheless, it  enables one, in principle,  to choose an anomaly 
free low-energy spectrum. In this study we include  
only a subset of fields which are relevant
for the generation of the $\mu$ and $m_{3}^{2}$ parameters
and only comment on the anomaly cancelation in our conclusions.
Also, in order to explore
a wider range of possibilities, the $U(1)_{\eta}$ assumption will be relaxed
in certain cases, and we will vary the $U(1)$$^{\prime}$ charge assignments.

The superpotential of the model is 
\begin{equation}
\label{w}
\begin{array}{cc}
W = & h_{s}S H_{1}  H_{2} + h_{u} u_{3}^c Q_{3}  H_{2} + h_{d} d_{3}^c Q_{3} 
 H_{1} + h_{e} e_{3}^c L_{3}  H_{1}\\
 & + h_{D} S D_{i} D^{c}_{i} + ({\rm self-couplings~of}~S,~ S^{\prime}).
\end{array}
\end{equation}
In Eqn.~(1) we include the usual Yukawa terms involving the third
generation fields, an effective $\mu$ term $h_{s}S H_{1}  H_{2}$, 
and indicate possible self-couplings of the singlet fields.

The soft supersymmetry breaking parameters are generated at the messenger
scale $\Lambda$ through gauge mediation. The breaking of the supersymmetry is
parameterized by the singlet $X$, whose scalar component acquires a vev
$\langle X \rangle$ and auxiliary component $F$ a vev $\langle F_{X} \rangle$.
The messenger fields $\Phi$ and $\bar{\Phi}$ are vector-like pairs
which transform under the SM
gauge group. Here, we assume only one such pair of $5$ and $\bar{5}$
of $SU(5)$. The superpotential term
\begin{equation}
W = \lambda X \Phi \bar{\Phi}, 
\end{equation}
where $\lambda$ is a Yukawa coupling, generates the supersymmetry
conserving and breaking
masses  for the messenger fields
$\sim \lambda\langle X\rangle$ and $\sqrt{|\lambda\langle F_{X}\rangle|}$, respectively.

Because $\Phi$ and $\bar{\Phi}$ are charged under
the SM gauge group, the effect of the supersymmetry breaking is propagated
at the quantum level to the observable sector via the usual gauge 
interactions. The messenger scale is
defined here as $\Lambda = \langle F_{X} \rangle / \langle X \rangle$. 
The gaugino masses are generated at one-loop \cite{DNS}
\begin{equation}
M_{i} = \frac{\alpha_{i}}{4 \pi}r_{i} \Lambda, 
\label{gaugino}
\end{equation} 
where $i=1^{\prime},~1,~2,~3$, and $r_{i}$ is the Dynkin index for $\Phi$ and
$\bar{\Phi}$. We choose, as is customary,  
a normalization such that the minimal
$5+\bar{5}$ model has $r_1=r_2=r_3=1$. In particular, we assume\footnote{
This assumption is formally inconsistent with an $E_{6}$ embedding
of the messengers $5 + \bar{5} \subset 27$, and is therefore
inconsistent with a true $E_{6}$ embedding of the models, which we do
not assume at this point.} 
(for simplicity only) that the messenger fields are not charged under
the additional $U(1)$, and hence,
$r_{1^{'}}=0$. The scalar masses arise at the two-loop level and are
given by 
\begin{equation}
m^{2}= 2 \Lambda^{2} \sum_{i} (\frac{\alpha_{i}}{4 \pi})^{2} C_{i}, 
\label{scalar}
\end{equation}     
where $C_{i}$ are the quadratic Casimirs of the observable sector gauge groups,
i.e., $C_{3}=4/3$ for $SU(3)_{c}$ triplets, $C_{2}=3/4$ for $SU(2)_{L}$
doublets, and $C_{1}=\frac{3}{5}Q_{Y}^{2}$ for the hypercharge. 
Again, the $U(1)^{\prime}$ does not 
contribute to the $m^{2}$ parameters since by assumption 
$\Phi$ and $\bar{\Phi}$ have zero $U(1)^{\prime}$ charges. 
We will take eqs.~(\ref{gaugino}) and (\ref{scalar})  
to be the boundary condition at a scale $\Lambda$, though more
generally one could choose a slightly different scale for the boundary.
(Our results do not depend on this assumption.)

The $A$ parameters arise only at the higher-loop order and  are very small. In our
study we take the $A$ parameters to be zero at the messenger
scale. Nevertheless, non-trivial values of the $A$ parameters arise 
from the one-loop renormalization-group evolution.
Assuming that the messenger fields $\Phi$ and $\bar{\Phi}$ are not
charged under the $U(1)$$^{\prime}$, the gaugino of the $U(1)$$^{\prime}$ and
the SM singlet fields are therefore massless at the
messenger scale\footnote{Even if $\Phi$ and $\bar{\Phi}$ are charged under
the $U(1)^{\prime}$, it will not affect the conclusions we have reached here,
since the contributions from the $U(1)^{\prime}$ to the soft mass parameters
are, in general,  small. It can affect, however, the (singlet) slepton
spectrum, which is otherwise given only by hypercharge loops.}.
However, $m_{S}^{2}$ acquires a non-zero value at the low energy scale  
due to loop corrections. In particular, it is driven rapidly to 
large and negative values 
due to the large couplings between $S$ and the exotic quark pairs $D$ and
$D^{c}$.   
Once $U(1)^{\prime}$ is broken, its gaugino and gauge boson are  degenerate in mass.

We use the renormalization group equations (RGE) to relate the
boundary conditions for the SSB parameters at the messenger scale to
their  values at  lower energies.
As the first step, the scale at which the additional $U(1)$$^{\prime}$ is
broken is determined and is required to be higher than the electroweak
scale (in order to be consistent with the experimental limits on the $Z^{\prime}$ gauge
boson). At this scale, the exotic quarks acquire heavy masses and
decouple from the theory. By
iterating this procedure, the gauge couplings at the messenger scale
$\Lambda = 10^{5}~{\rm GeV}$ are determined from their 
well-known electroweak-scale values (taking into account the
contributions of the exotic matter).
At the second step,
the values of $\mu$ (or equivalently, $h_{s}$)
and $\tan{\beta} = \langle H_{2}\rangle / \langle H_{1}\rangle$ at the
minimum of the Higgs potential are determined, the former by
fixing the mass of the $Z$ boson 
and the latter by using the result for $\mu$ and $m_{3}^{2} = A_{s}\mu$.
Given $\tan\beta$, the Yukawa couplings at the electroweak scale are determined
from the respective fermion masses. The procedure is then iterated
in order to determine the correct Yukawa couplings for the
$t$ and $b$ quarks and the $\tau$-lepton. 
The desired 
solution for the minimum of the Higgs
potential which correctly reproduces the $Z$, $t$, $b$ and $\tau$
masses is found and the $Z^{\prime}$ mass and mixing, in addition to
the sparticle spectrum, are predicted.
Given our assumptions,
the free parameters in the analysis are $\Lambda$, the number $n_{D}$ of $D$,
$D^{c}$ pairs that couple to $S$, the corresponding  Yukawa couplings
$h_{D}$, and (in the case of two SM singlets) the singlet self-coupling. 
The product of $n_{D}$ and $h_{D}$ is constrained by electroweak
breaking and also by requiring a sufficiently heavy $Z^{\prime}$.
Below we fix $n_{D} = 3$ and use different values of $h_{D} =
{\cal{O}}(1)$. We comment on the variation of the parameters
at the end of the next  section.

\section{The radiatively induced Higgs mass parameters}

The Higgs potential contains three contributions
\begin{equation}
V=V_{F}+V_{D}+V_{soft}.
\label{Vtot}
\end{equation}
In the following two subsections, we study two cases with one or
two SM singlet fields in the model. In the second case, the
additional singlet $S^{\prime}$ has a coupling $S S^{'2}$ allowed by gauge
invariance. The most general renormalizable superpotential involving the Higgs
and two singlet fields is
\begin{equation}
\label{wcomplete} 
W = h_{s} S H_{1}  H_{2} + h_{s'}S S^{'2}. 
\end{equation} 
The potential (\ref{Vtot}) is given in this case  by  
\begin{equation}
\label{vf}
V_{F} = |h_{s} H_{1} H_{2}+h_{s'}S^{'2}|^{2}+
|h_{s}S|^{2} (|H_{1}|^{2}+|H_{2}|^{2})+ 4|h_{s'}SS^{\prime}|^2;  
\end{equation} 
\begin{equation}
\label{vd}
V_{D} = \frac{G^{2}}{8}(|H_{1}|^{2}-|H_{2}|^{2})^{2} +
\frac{g_{2}^{2}}{2}|H_{1}^{\dagger}H_{2}|^{2} +
\frac{g_{1^{\prime}}^2}{2}(Q_{1}|H_{1}|^{2}+Q_{2}|H_{2}|^{2}+Q_{S}|S|^{2}+Q_{S^{\prime}}|S^{\prime}|^{2})^{2}; 
\end{equation}
\begin{equation}
\label{vsoft}
\begin{array}{cc} 
V_{soft} &= m_{1}^{2}|H_{1}|^{2} + m_{2}^{2}|H_{2}|^{2} + m_{S}^{2}|S|^{2}
+ m_{S^{\prime}}^{2}|S^{\prime}|^{2} \\
 & + ( A_{s} h_{s} S H_{1}  H_{2} + {\rm h.c.}) +
(A_{s'} h_{s'} S S^{'2} + {\rm h.c.}),  
\end{array}
\end{equation} 
where $G^{2}=g_{Y}^{2}+g_{2}^{2}$. 
The one singlet case is given by simply setting $h_{s^{\prime}} = 0$ and $Q_{S^{\prime}}= 0$.

\subsection{Radiatively breaking the $U(1)$$^{\prime}$ symmetry}

The experimental constraint on the mass of the $Z^{\prime}$ 
can be satisfied if 
the $U(1)$$^{\prime}$ is broken at the TeV scale, 
which requires $\langle H_{1} \rangle, \langle 
H_{2} \rangle \ll \langle S \rangle,\,\langle S^{\prime}
\rangle$. This separation is indeed realized in our examples and
the determination of the $S$ vev can therefore be separated 
to a very good approximation from that of 
the Higgs vev's. We will  illustrate the radiative
breaking of the $U(1)$$^{\prime}$ symmetry in the case of a single 
SM singlet $S$.
The scalar potential for $S$ reads 
\begin{equation} 
\label{vs}
V= m_{S}^{2}|S|^{2} + \frac{g_{1^{\prime}}^2}{2} (Q_{S}|S|^2)^{2}. 
\end{equation} 
It acquires a vev  $\langle S \rangle = s/\sqrt{2}$ where
\begin{equation} 
\label{s}
s^{2}= - \frac{2 m_{S}^{2}}{g_{1^{\prime}}^{2}Q_{S}^{2}}, 
\end{equation} 
if the evolution of $m_{S}^{2}$ can be neglected near the minimum. Hence, a
large value for $s$ occurs for
$m_{S}^{2}$ large and negative. This is achieved by the
order unity Yukawa couplings between $S$ and exotic quark pairs $D$
and $D^{c}$ (with scalar mass-squares $m_{D,\,D^{c}}^{2}(\Lambda) \gg
m_{S}^{2}({\Lambda}) \simeq 0$), 
which rapidly diminish $m_{S}^{2}(Q < \Lambda)$ via the usual
renormalization group evolution.  
The mass of the $Z^{\prime}$ boson, which is 
independent of $g_{1^{\prime}}$, is
\begin{equation}
M_{Z^{\prime}} \sim g_{1^{\prime}} Q_{S} s \sim \sqrt{2|m_{S}^{2}|},
\end{equation} 
with the $Z-Z^{\prime}$ mixing angle $\alpha_{Z-Z^{\prime}} = {\cal O}
(M_{Z}^{2}/M_{Z^{\prime}}^{2})$.   The $Z^{\prime}$ mass 
and the $U(1)$$^{\prime}$ scale are  determined by
the only scale in the problem, $\Lambda$ (which is encoded in $m_{S}^{2}$). 
 
\subsection{One singlet models}

We first consider models with only one singlet field, with its
$U(1)$$^{\prime}$ charge satisfying $Q_S+Q_{H_{1}}+Q_{H_{2}}=0$. 
The superpotential
is given by the first five terms in (\ref{w}).   
The vev of $S$ generates an effective $\mu$ parameter
$\mu=h_{s}s/\sqrt{2}$. The $A$-term associated with $S H_{1}  H_{2}$,
which is non-zero at the electroweak scale due to loop corrections, 
generates an effective $m_{3}^{2}$ for the two Higgs doublets 
$m_{3}^{2}= A_{s}\mu$. 
In addition, the $U(1)$$^{\prime}$ $D$-term generates corrections to the Higgs
scalar masses $\delta m_{1,\,2}^{2} = \frac{g_{1^{\prime}}^2}{2}Q_{1,\,2} Q_{S} s^2$. 
Defining $\langle H_{1}^0 \rangle = v_1/\sqrt{2}$ and $\langle H_{2}^0 \rangle =
v_2/\sqrt{2}$, the Higgs potential for $v_{1,2}$ at the minimum for $s$ is 
\begin{equation}
\begin{array}{cc}
\label{vew}
V=& \frac{\mu^{2}}{2} (v_1^2 + v_2^2) + \frac{G^2}{32}(v_2^{2}-v_{1}^{2})^{2} + 
\frac{1}{2}(m_{1}^{2} + \delta m_{1}^{2})v_1^{2} \\
 & +  \frac{1}{2}(m_{2}^{2} +
\delta m_{2}^{2})v_2^{2} - m_3^2 v_1 v_2 + \cdots , 
\end{array}
\end{equation} 
where we use a suitable gauge rotation to render $v_{1}$ and $v_{2}$ real and
positive, and we neglect small corrections from the $U(1)$$^{\prime}$ $D$ term
which are quadratic in $v_{1}/v_{2}$.
(The usual MSSM loop corrections can also be absorbed in $\delta m_{i}^{2}$,
but are a secondary effect at this level.) 
We present in the following
two numerical examples with different choices of $Q_{1}$ and $Q_{2}$.

In the first example, we choose the $U(1)_{\eta}$ assignments $Q_{1} = 1$, $Q_{2} =
4$ and $Q_{S}=-Q_{1}-Q_{2}$. 
The initial and final values of the 
parameters are listed in Table \ref{oneS1}. We take 
$\Lambda = 10^5$ GeV and $3$ pairs of exotic quark singlets. 

\begin{table}
\begin{center}
\begin{tabular}{|c|c|c|c|c|c|}   

  & $\Lambda$  & $M_{Z}$ & $$  & $\Lambda$ & $M_{Z}$  \\ \hline   
$h_{u}$ & $0.84$ & $0.98$ & $h_{d}$ & $0.30 $ & $0.42 $ \\ \hline
$h_{e}$ & $0.17$ & $0.18$ & $h_{D}$ & $0.70 $ & $0.84 $ \\ \hline
$h_{s}$ & $0.47$ & $0.40$ & $A_{s}~({\rm GeV})$ & $0 $ & $-53 $ \\ \hline
$A_{u}~({\rm GeV})$ & $ 0$ & $449$ & $A_{d}~({\rm GeV})$ & $0 $ & $573 $ \\
\hline  
$A_{e}~({\rm GeV})$ & $ 0$ & $24$ & $A_{D}~({\rm GeV})$ & $0 $ & $494 $ \\
\hline 
$m_{1}^{2}~({\rm GeV})^2$ & $(350)^{2}$ & $(83)^{2} $ & 
             $m_{2}^{2}~({\rm GeV})^2$ & $(350)^{2} $ & $-(778)^{2} $ \\ \hline
$m_{S}^{2}~({\rm GeV})^2$ & $0$ & $-(821)^{2}$ & 
             $m_{D^{(c)}}^{2}~({\rm GeV})^2$ & $(1310)^{2}$ & $(1440)^{2}$ 

\end{tabular}
\caption{The initial and final values for the first example of the one singlet
case. $n_{D}=3$ and $\Lambda=10^{5}~{\rm GeV}$. }
\label{oneS1}
\end{center}
\end{table}

The vev of the singlet is $s=3720~{\rm GeV}$, the vev's of the Higgs doublets are
$v_{1}=14~{\rm GeV}$ and $v_2= 245~{\rm GeV}$, resulting in a solution with $\mu =
1050~{\rm GeV}$ and $\tan{\beta}=18$. The effective $m_{3}^2$ is $\sim
(235\,{\text GeV})^2$. The $Z^{\prime}$ mass is $M_{Z^{\prime}}=1110$ GeV  and  
the $Z-Z^{\prime}$ mixing angle is $\alpha_{Z-Z^{\prime}} =
0.004$. The  (tree-level) spectrum of the
CP even physical Higgs is $m_{h_{1}}=124~{\rm GeV}$, $m_{h_{2}}=995~{\rm GeV}$,
$m_{h_{3}}=1090~{\rm GeV}$, while $m_{h_{1}}= 154~{\rm GeV}$ at one loop
(with negligible corrections to $m_{h_{2,\,3}}$).
The CP odd Higgs scalar 
and the charged Higgs masses are $m_{A} \simeq m_{H^{\pm}}=993~{\rm GeV}$. The heaviest
CP even Higgs scalar $h_{3}$ is mainly composed of the singlet $S$, associated with the
breaking of the $U(1)$$^{\prime}$. The second heaviest CP even Higgs, 
the CP odd
Higgs and the charged Higgs fields form the $SU(2)$ doublet that is not
associated with the ${SU(2)}\times {U(1)}_Y$ breaking.   

 The masses of the two charginos are $m_{\tilde{\chi}_{1}^{\pm}}=266~{\rm GeV}$ and
 $m_{\tilde{\chi}_{2}^{\pm}}=1060~{\rm GeV}$. The lightest (heaviest) chargino is
 predominantly a gaugino (Higgsino). The spectrum of the neutralinos is
 $m_{\tilde{\chi}_{1}^{0}}=142~{\rm GeV}$, $m_{\tilde{\chi}_{2}^{0}}=266~{\rm GeV}$, 
 $m_{\tilde{\chi}_{3}^{0}}=1060~{\rm GeV}$, $m_{\tilde{\chi}_{4}^{0}}=1060~{\rm GeV}$, 
 $m_{\tilde{\chi}_{5}^{0}}=1120~{\rm GeV}$, $m_{\tilde{\chi}_{6}^{0}}=1120~{\rm GeV}$. 
 In the limit of neglecting $v_1$ and $v_2$, the two lightest
 neutralinos are just $\tilde{B}$ and $\tilde{W_3}$, i.e., the 
Bino and the Wino. $\tilde{\chi}_{3,\,4}$ are linear combinations of Higgsinos
with 
nearly degenerate masses $\sim \mu=h_s s/\sqrt{2}$; and  $\tilde{\chi}_{5,\,6}$
are linear combinations of the other gaugino 
$\tilde{B}^{\prime}$ and the singletino $\tilde{S}$ with degenerate masses $\sim
M_{Z^{\prime}}$.

In the second example, we consider a special case in which the $U(1)$$^{\prime}$ charge
of $H_{2}$ is set to be zero, i.e., $Q_{1}=-Q_S=5$. In this case,
there is no 
$U(1)$$^{\prime}$ $D$-term correction to the $H_{2}$ mass squared, reducing the value of $\tan{\beta}$ at the minimum of the potential. We
present the numerical values of the parameters chosen for this example in Table
\ref{oneS2}. Again, $\Lambda$ is chosen to be $10^{5}~{\rm GeV}$ and $n_{D}=3$. 

\begin{table}
\begin{center}
\begin{tabular}{|c|c|c|c|c|c|}   

  & $\Lambda$  & $M_{Z}$ & $$  & $\Lambda$ & $M_{Z}$  \\ \hline   
$h_{u}$ & $0.88$ & $1.0$ & $h_{d}$ & $0.06 $ & $0.09 $ \\ \hline
$h_{e}$ & $0.04$ & $0.04$ & $h_{D}$ & $0.70 $ & $0.85 $ \\ \hline
$h_{s}$ & $0.36$ & $0.31$ & $A_{s}~({\rm GeV})$ & $0 $ & $-48 $ \\ \hline
$A_{u}~({\rm GeV})$ & $ 0$ & $443$ & $A_{d}~({\rm GeV})$ & $0 $ & $555 $ \\
\hline  
$A_{e}~({\rm GeV})$ & $ 0$ & $34$ & $A_{D}~({\rm GeV})$ & $0 $ & $494 $ \\
\hline 
$m_{1}^{2}~({\rm GeV})^2$ & $(351)^{2}$ & $(363)^{2} $ & 
             $m_{2}^{2}~({\rm GeV})^2$ & $(351)^{2} $ & $-(815)^{2} $ \\ \hline
$m_{S}^{2}~({\rm GeV})^2$ & $0$ & $-(819)^{2}$ &   
             $m_{D^{(c)}}^{2}~({\rm GeV})^2$ & $(1310)^{2}$ & $(1440)^{2}$ 
\end{tabular}
\caption{The initial and final values for the second example of the one singlet
case. $n_{D}=3$ and $\Lambda=10^{5}~{\rm GeV}$. }
\label{oneS2}
\end{center}
\end{table}

The vev of the singlet is $s=3800~{\rm GeV}$, the vev's of the Higgs doublets are
$v_{1}=60~{\rm GeV}$ and $v_2= 238~{\rm GeV}$, resulting in a solution with $\mu =
818~{\rm GeV}$ and $\tan{\beta}=4$, and thus $m_3^2=(198\,{\text GeV})^{2}$. In this
case, $M_{Z^{\prime}}=1140$ GeV and $\alpha_{Z-Z^{\prime}} = 3.1 \times 10^{-4}$. The masses of the
CP even physical Higgs are $m_{h_{1}}=157~{\rm GeV}$, $m_{h_{2}}=434~{\rm GeV}$,
$m_{h_{3}}=1130~{\rm GeV}$ at tree level, while $m_{h_{1}}= 181~{\rm GeV}$ at one loop. 
The CP odd Higgs scalar 
and the charged Higgs masses are $m_{A}=409~{\rm GeV}$ and 
$m_{H^{\pm}}=413~{\rm GeV}$, respectively. 
The pattern of the
spectrum is similar to that given in the previous example. 

The masses of the charginos and neutralinos in this example are
$m_{\tilde{\chi}_{1}^{\pm}}=262~{\rm GeV}$ and 
$m_{\tilde{\chi}_{2}^{\pm}}=843~{\rm GeV}$;  $m_{\tilde{\chi}_{1}^{0}}=140~{\rm GeV}$,
$m_{\tilde{\chi}_{2}^{0}}=262~{\rm GeV}$,  
$m_{\tilde{\chi}_{3}^{0}}=834~{\rm GeV}$, $m_{\tilde{\chi}_{4}^{0}}=841~{\rm GeV}$, 
$m_{\tilde{\chi}_{5}^{0}}=1130~{\rm GeV}$ 
and $m_{\tilde{\chi}_{6}^{0}}=1140~{\rm GeV}$. It again
falls into the pattern of the chargino and neutralino spectrum discussed in the
previous example. 

In both examples squark and gluino masses are in the $1200 - 1400$ GeV
range. The next to lightest sparticle (NLSP) is the lightest neutralino, which 
is predominantly the bino, i.e., the gaugino of the $U(1)_Y$.
The evolution of the parameters is illustrated in Figs.~1 and 2.

\begin{figure}
\centerline{
\hbox{
\epsfxsize=2.8truein
\epsfbox[70 32 545 740]{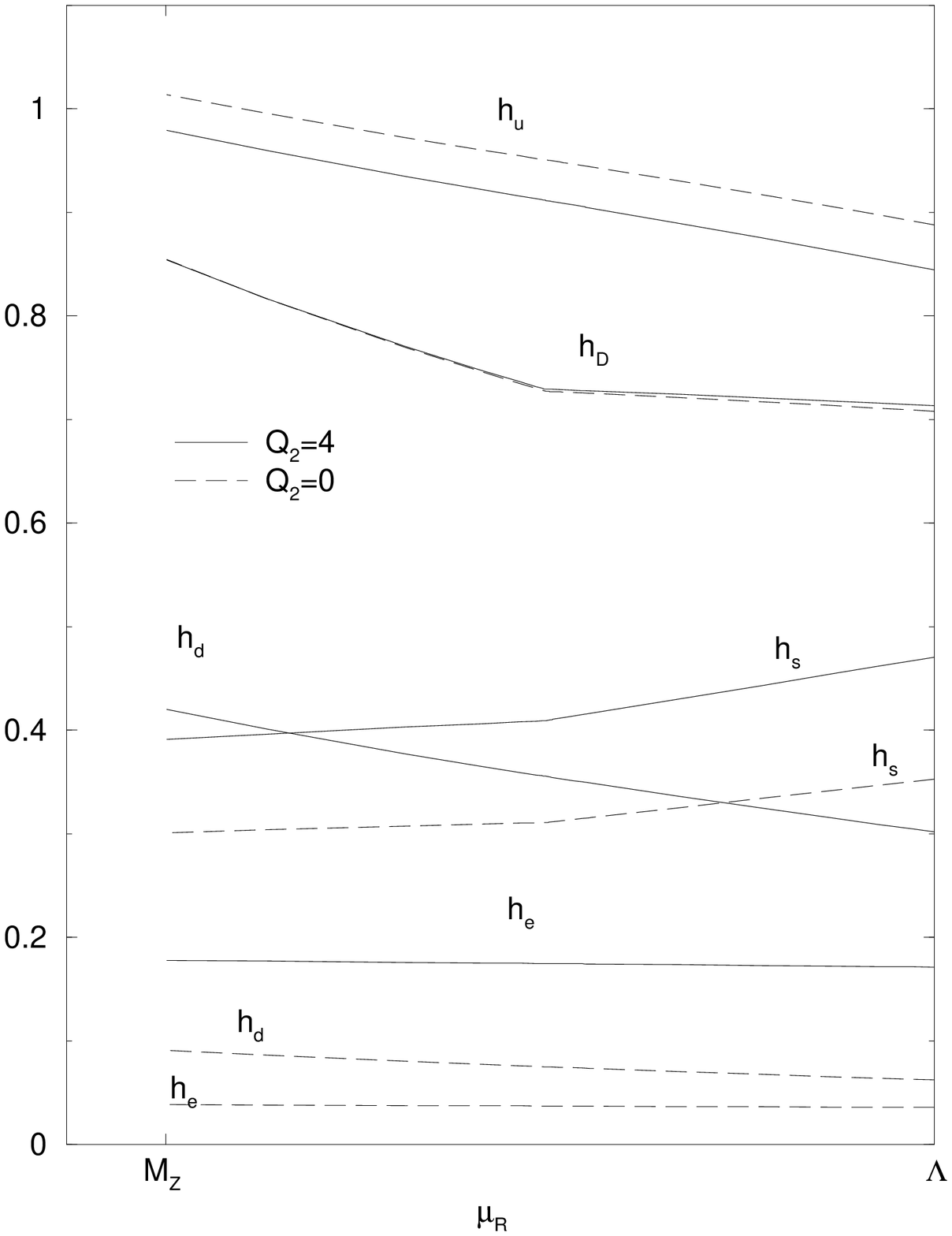}
\hskip 0.25truein
\epsfxsize=2.8truein
\epsfbox[70 32 545 740]{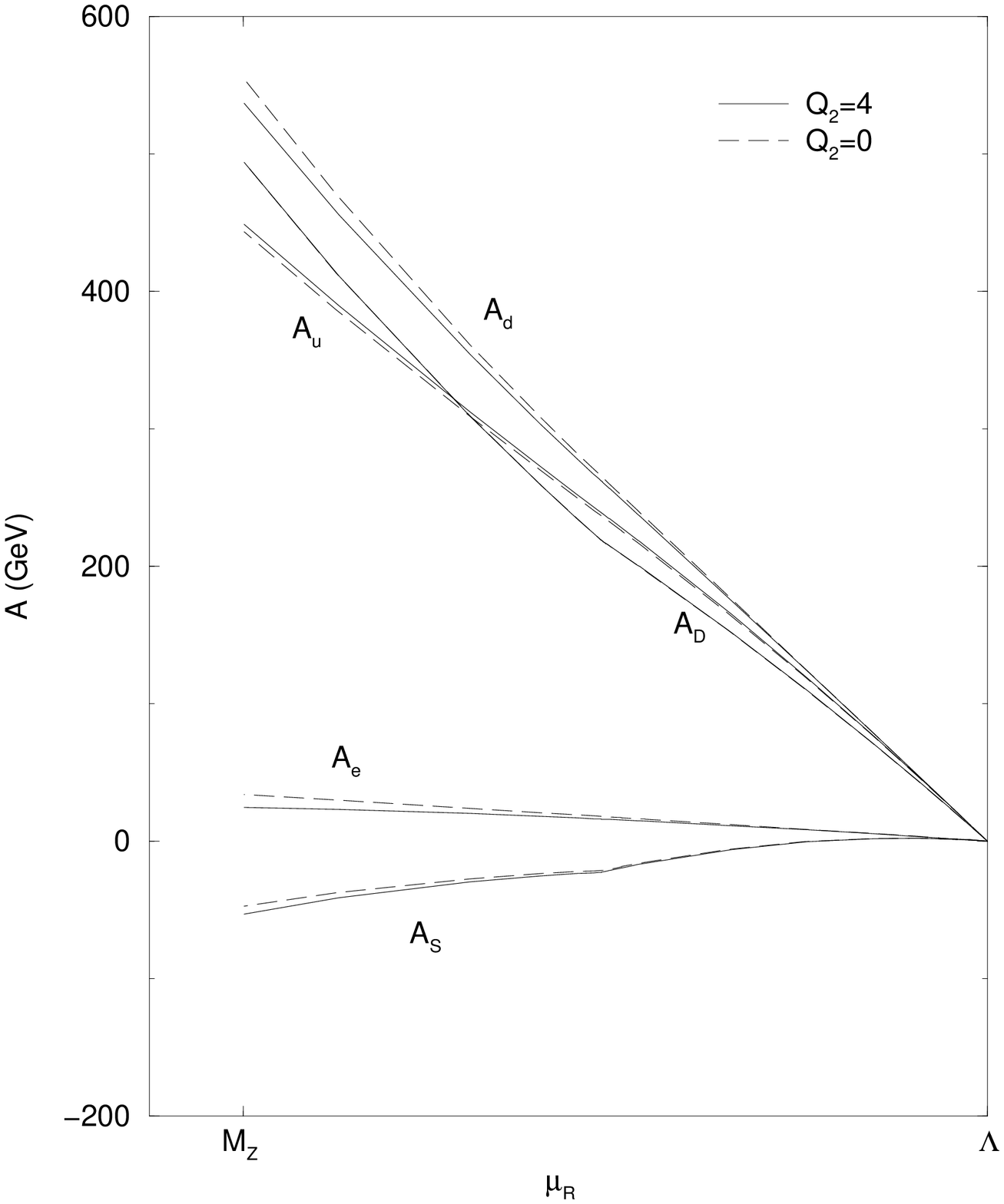}
}
}
\caption{
The RGE evolution of the Yukawa and trilinear couplings in the two
examples of the one singlet case. 
$\mu_{R}$ is the renormalization scale and $\Lambda =10^{5}$ GeV.   
}
\vskip 0.1truein
\centerline{
\hbox{
\epsfxsize=2.8truein
\epsfbox[70 32 545 740]{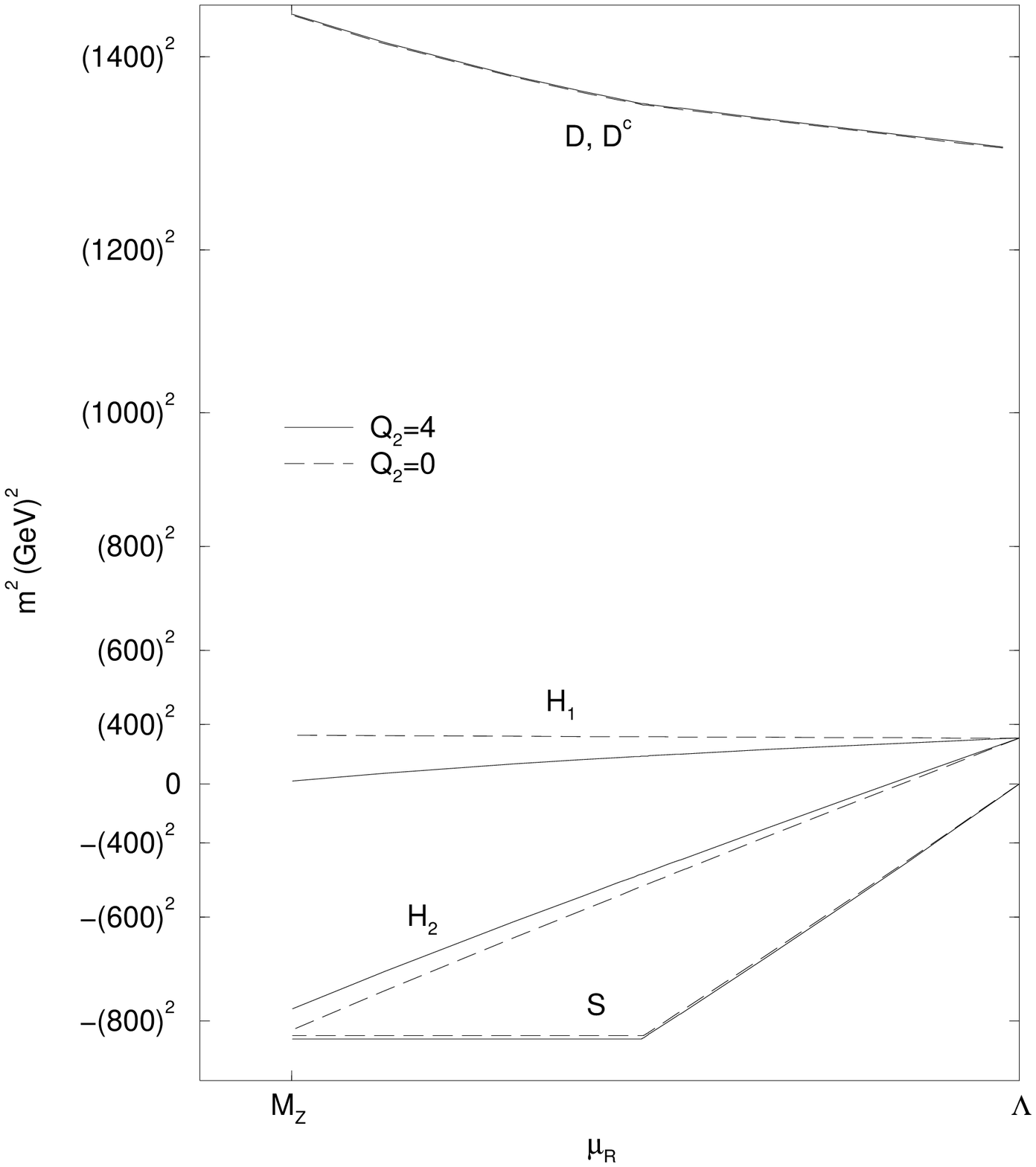}
}
}
\caption{
The RGE evolution of the mass squared parameters in the two examples of the one singlet case. 
}
\end{figure}

\subsection{Multi-singlet models}

We now consider the case with two singlets fields $S$ and $S^{\prime}$. Their
charges are such that the coupling $S S^{'2}$ is allowed by gauge invariance. 
The $F$ terms of the scalar potential (\ref{vf}) indicate that in addition
to the 
effective $\mu$ term, $\mu=h_{s} \langle S \rangle$, there is an additional contribution to 
$m_{3}^{2}$ arising from the mixed term between $H_{1}  H_{2}$ and
$S^{'2}$. Hence, the total effective $m_{3}^{2}$ is given by 
\begin{equation}
m_{3}^{2} = h_{s} h_{s'} \langle S^{\prime} \rangle ^{2} +  A_{s} h_{s} \langle S 
\rangle, 
\end{equation} 
when both $S$ and $S^{\prime}$ acquire non-zero vev's. There are also corrections
to the $H_{1}$ and $H_{2}$ mass-squared parameters from the $D$-term eqn.~(\ref{vd}), 
$\delta m_{1/2}^{2} = \frac{g_{1^{\prime}}^2}{2}Q_{1/2} (Q_{S} s^2 + Q_{S^{\prime}}
s^{'2})$, where $s^{(')}= \sqrt{2} \langle S^{(')} \rangle$ are the
vev's of $S$ and $S^{\prime}$. The radiative $U(1)$$^{\prime}$ symmetry breaking is again
achieved by the coupling between $S$ and the exotic quark pairs $D$ and
$D^{c}$. In this scenario, the self-coupling between $S$ and $S^{\prime}$ 
stabilizes the vacuum of the two-singlet potential. To
satisfy the phenomenological constraint on $M_{Z^{\prime}}$, i.e., 
to ensure that the $U(1)$$^{\prime}$
is  broken at $\sim {\cal O} ({\rm TeV})$, the Yukawa coupling
$h_{s^{\prime}}$ has to be small. Defining the vev's of the Higgs doublets as in the
previous cases, the Higgs potential at the minimum again takes the form
eqn. (\ref{vew}).

As an explicit example, we choose the charges of the Higgs doublets and the singlets
to be $Q_{1} = 1$, $Q_{2} = 4$, $Q_{S}=-Q_{1}-Q_{2}$ and
$Q_{S^{\prime}}= - Q_{S}/2$.
We again choose $\Lambda=10^{5}~{\rm GeV}$ and $n_{D}=3$.  
We list the initial and final values of the relevant parameters in Table \ref{twoS}.

\begin{table}
\begin{center}
\begin{tabular}{|c|c|c|c|c|c|}   

  & $\Lambda$  & $M_{Z}$ & $$  & $\Lambda$ & $M_{Z}$  \\ \hline   
$h_{u}$ & $0.84$ & $0.98$ & $h_{d}$ & $0.18 $ & $0.26 $ \\ \hline
$h_{e}$ & $0.10$ & $0.11$ & $h_{D}$ & $1.00 $ & $1.13 $ \\ \hline
$h_{s}$ & $0.45$ & $0.36$ & $h_{s^{\prime}}$ & $0.12$ & $0.10 $ \\ \hline
$A_{s}~({\rm GeV})$ & $0 $ & $-50 $ & $A_{s^{\prime}}~({\rm GeV})$ & $0 $ & $-29 $ \\ \hline
$A_{u}~({\rm GeV})$ & $ 0$ & $442$ & $A_{d}~({\rm GeV})$ & $0 $ & $538 $ \\
\hline  
$A_{e}~({\rm GeV})$ & $ 0$ & $31$ & $A_{D}~({\rm GeV})$ & $0 $ & $477 $ \\
\hline 
$m_{1}^{2}~({\rm GeV})^2$ & $(352)^{2}$ & $(306)^{2} $ & 
             $m_{2}^{2}~({\rm GeV})^2$ & $(352)^{2} $ & $-(758)^{2} $ \\ \hline
$m_{S}^{2}~({\rm GeV})^2$ & $0$ & $-(966)^{2}$ & 
             $m_{S^{\prime}}^{2}~({\rm GeV})^2$ & $0$ & $(15)^{2}$  \\ \hline 
$m_{D^{(c)}}^{2}~({\rm GeV})^2$ & $(1280)^{2}$ & $(1400)^{2}$ 
              &  &  &
\end{tabular}
\caption{The initial and final values for the example of the two singlet
case. $n_{D}=3$ and $\Lambda=10^{5}~{\rm GeV}$. }
\label{twoS}
\end{center}
\end{table}

The vev's of the singlets are $s=4430~{\rm GeV}$ and $s^{\prime}=1760$, and
those of the Higgs doublets are
$v_{1}=23~{\rm GeV}$ and $v_2= 245~{\rm GeV}$, resulting in a solution with
$\mu = 1110~{\rm GeV}$ and $\tan{\beta}=10$, which is accidentally similar to
the solution in our first example. 
The effective $m_{3}^2=(407\,{\text GeV})^2$. The
$Z^{\prime}$ mass is $1380$ GeV, with the $Z-Z^{\prime}$ mixing angle
$0.004$. The masses
for the four CP even scalars, which are mixtures of the Higgs scalars and the
scalar components from the singlets $S$ and $S^{\prime}$ are $m_{h_{1}}=124~{\rm GeV}$,
$m_{h_{2}}=459~{\rm GeV}$, $m_{h_{3}}=1090~{\rm GeV}$ and
$m_{h_{4}}=1390~{\rm GeV}$, with $m_{h_{1}}= 154~{\rm GeV}$ at one loop. 
The two CP 
odd scalar masses are $m_{A_1}=138~{\rm GeV}$ and $m_{A_2}=1080~{\rm GeV}$. 
The charged Higgs scalar mass is $m_{H^{\pm}}=1090~ {\rm GeV}$. 
Therefore, $h_{3},~A_{2}$ and 
$H^{\pm}$ approximately form an $SU(2)$ doublet which is not involved in the
electroweak breaking. 
$A_1$ is predominantly associated with the two singlets.
Its lightness is readily understood in terms of the 
small values of $h_{s^{\prime}}$ and $A_{s^{\prime}}$, and 
the extra global $U(1)$ symmetry which occurs for 
$h_{s^{\prime}} = 0$ and $A_{s^{\prime}} = 0 $.
$h_4$ is also mostly associated with the two singlets.  
As noted above, the $\mu$ parameter in this  and the first example 
accidentally has similar values. However, the initial value of
$h_{D}$ has to be increased in this example 
($1$ compared to $0.7$ in the previous case)
to counteract the effect of the self-coupling between the two singlets, so
that the singlet vev's could both have large values.

The masses of the two charginos are $m_{\tilde{\chi}_{1}^{\pm}}=266~{\rm GeV}$ and
$m_{\tilde{\chi}_{2}^{\pm}}=1130~{\rm GeV}$; the masses of the seven neutralinos in
this example are $m_{\tilde{\chi}_{1}^{0}}=142~{\rm GeV}$,
$m_{\tilde{\chi}_{2}^{0}}=266~{\rm GeV}$,  
$m_{\tilde{\chi}_{3}^{0}}=348~{\rm GeV}$, $m_{\tilde{\chi}_{4}^{0}}=1130~{\rm GeV}$, 
$m_{\tilde{\chi}_{5}^{0}}=1130~{\rm GeV}$, $m_{\tilde{\chi}_{6}^{0}}=1300~{\rm GeV}$ and
$m_{\tilde{\chi}_{7}^{0}}=1470~{\rm GeV}$. It approximately follows the 
pattern of the chargino, neutralino spectrum discussed in the
previous examples, which follows from  limit of $v_1, v_2 \ll s
(s^{\prime})$ and the similar values of $\mu$. However,   
$\tilde{\chi}_{3}^{0}$, coming from the self-coupling between the two
singlet fields with mass $\sim h_{s^{\prime}} s$, is a new feature.
Squark and gluino masses are again in the $1200 - 1400$ GeV
range and the NLSP is again the lightest neutralino.

\subsection{Variation of the parameters}
\label{sec:variation}

As we discussed in Sec. II, the free parameters in our analysis are $\Lambda$, 
$h_{D}$ and $n_{D}$. In our numerical examples, we choose $\Lambda=10^5$
GeV. If $\Lambda$ is varied while the other parameters are kept fixed, the
scale at which the $U(1)$$^{\prime}$ is broken  changes 
so that $\Lambda^{\prime}/\Lambda$ 
is approximately a constant. For example, if we raise
 $\Lambda$ to be $10^6$ GeV in the first example, the singlet vev becomes
$s \simeq 33400$ GeV. 
This is because RSB depends on the evolution interval, and not
on the actual location of the boundary.

Next, consider $n_D$, the number of the exotic quarks that couple to the
singlet field $S$ with coefficient $h_D$. 
Choosing a smaller $n_D < 3$ and keeping  $\Lambda$ and $h_D$ fixed, the
$U(1)$$^{\prime}$
breaking scale is reduced, and the associated $Z^{\prime}$ mass is diminished.
For example,  setting  $n_D=2$ in the first example instead of $3$, 
the singlet vev is now  $s = 3080$ GeV, resulting in
$M_{Z^{\prime}}=919$ GeV with a mixing angle $\alpha_{Z-Z^{'}}=0.006$.  
In particular, with $n_D=1$, $h_D=0.7$ is too small
to generate a solution with $M_{Z^{\prime}} > 700$ GeV
(which is a model dependent experimental lower bound). On the other hand, if $n_D$ is
fixed, increasing/decreasing 
$h_D$ will raise/lower the $U(1)$$^{\prime}$ scale
and solutions for $n_{D} = 1$ exist for larger values of $h_{D}$. 
Hence, it is the product of $h_D$
and $n_D$ which is constrained 
by the  $Z^{\prime}$ mass. Note that an upper bound on the Yukawa
couplings  $h_D \simeq {\cal{O}}(1)$
exists only if one requires
the model to still be perturbative at some high energy scale.        
Also, the variation with $n_{D}$ discussed above is similar for different
exotic quark quantum numbers
(for example, for two pairs of exotic quark singlets
instead of  an exotic quark doublet pair).

In our numerical analysis, we have identified, for simplicity, the messenger scale and the
scale parameter $\Lambda$
which determines the boundary conditions for the
soft mass parameters eqs.~(\ref{gaugino}) and (\ref{scalar}).
Differentiating these two scales will not change our conclusions.
For example, raising 
the boundary scale for the RGE evolution in the first example to 
$5\times 10^5$ GeV, while keeping $\Lambda=10^{5}$ GeV in
eqs.~(\ref{gaugino}) and (\ref{scalar})
leads to a solution with a larger singlet vev $s = 4430$ GeV 
due to the longer evolution interval. The $A_s$ parameter is also
larger in this case, $A_s =-86$ GeV, which results in a slightly smaller
$\tan{\beta}=13$ at the electroweak scale.

\section{Summary and conclusions}

It  has been shown that a $U(1)$$^{\prime}$ scale may be
generated naturally and radiatively one or two orders of
magnitude below the messenger scale. Upon integrating out the  
$U(1)$$^{\prime}$ sector, the supersymmetry conserving ($\mu$) and
breaking ($m_{3}^{2}$) dimensionful Higgs mixing parameters
are generated, resolving the $\mu$ problem in the otherwise
attractive class of gauge-mediation models. 
A natural consequence of the models with only one singlet is that
no new physical phases appear in the soft parameters, eliminating
potentially unacceptable contributions to CP violating amplitudes
which are flavor conserving and generically persist in gauge-mediation
models. Other implications include
new contributions to the effective quartic coupling
which lift the lightest Higgs boson mass to $m_{h} \gtrsim 150$ GeV. 
The $U(1)$$^{\prime}$ dynamics also adds to the already strong
predictive power of the gauge-mediation framework, as the scalar,
fermion and vector electroweak sectors are extended and new exotic
matter is predicted at a few TeV scale. 
The Higgs, neutralino and chargino sectors are now extended and contain new
degrees of freedom, but no new light pseudo Goldstone bosons. The
additional gauge boson typically has a ${\cal{O}}$(TeV) mass and negligible
mixing with the $Z$-boson. Its exact mass depends on the number of
exotic quarks and on the strength of their Yukawa couplings.

We find the usual near equality between $|\mu|$ and the 
gluino mass that often appears 
in various variants of the MSSM.
The gluino is heavy, which is a  generic prediction of the gauge-mediation
framework, and hence it naturally leads to a relatively
large value of $\mu$.
At the same time it implies that electroweak symmetry
breaking exhibits in our case the typical gauge
mediation tuning $\sim |M_{Z}/M_{\mbox{\tiny gluino}}| \sim 1/10$, 
rather than a new tuning due to  the $U(1)$$^{\prime}$
dynamics. It is worth stressing that the Higgs
mixing parameter in the scalar potential $m_{3}^{2} = \mu A_{s}$ 
(in the one singlet case) is a
geometrical mean of the superpotential Higgs mixing parameter $\mu$
and a radiatively generated (small) trilinear coupling $A_{s}$. Since
$\mu$ is proportional to a large vev
(and the  heavy gluino implies further that $|\mu|$ is not suppressed by a
small coupling) the geometrical mean $m_{3}
\sim {\rm a}\,\,{\rm few}\,\times\, 100$ GeV is sufficiently large.

While we have focused on the low-energy aspects of this one-scale
model, it is interesting to consider its embedding in a high-energy
theory, and in particular a unified or string theory. Unification of
gauge couplings at Planckian energies constrains the exotic matter
which is charged under the SM to fall into complete multiplets of a
unified group, e.g., $D + D^{c} \rightarrow 5 + \bar{5}$ of $SU(5)$.
It further constrains the number of such extra pairs of multiplets,
e.g., $n_{D} =\ n_{5} \lesssim 4$ \cite{MRK}. The counting now includes also the messenger
multiplets but may be modified by various considerations such as 
the exact mass (and hence, messenger) scale,
the presence of large Yukawa couplings (as in our case) 
and the normalization of the $U(1)$ factor(s). 
Anomaly cancelation suggests that the SM, exotic
multiplets  (and messengers - if charged under $U(1)^{\prime}$) 
are embedded in $27$ multiplets of E$_{6}$. The most straightforward 
realization which is consistent with the anomaly constraints 
(but which does not coincide with a standard $E_{6}$ model) is
that only the third family carries $U(1)$$^{\prime}$ charges, and the
extra $5 + \bar{5}$ that are required for the anomaly cancelation
correspond to one pair of exotics and to the usual Higgs doublets.
The messengers are also not charged under the $U(1)$$^{\prime}$ in this case.
We have shown that such a model is consistent 
(for a large  Yukawa coupling $h_{D} \sim 1$) but generically predicts a
lighter $Z^{\prime}$. 
This simple model, however, does not unify.
Unified (and hence, automatically anomaly-free) 
models require a more complicated embedding,
typically contain more than
one pair of exotic quarks, and further
require some separation mechanism between the low-energy 
Higgs doublets and the
heavier exotic leptons which appear. 
Such an embedding, however, is possible.
An explicit example which satisfies all constraints (including
unification of couplings) was given in Ref.~\cite{LW}.

Another issue that may arise is the kinetic mixing between $U(1)_Y$ and 
$U(1)^{\prime}$. If at the fundamental scale, the mixed trace between
$U(1)_Y$ and $U(1)^{\prime}$ does not vanish, a term in the Lagrangian that
mixes the field strength of the two $U(1)$'s could arise through loop
corrections at the low energy scale. A non-orthogonal transformation of
the gauge fields associated with the two $U(1)$'s is required so that the
kinetic energy terms can be written in their canonical forms. This
transformation effectively shifts the $U(1)^{\prime}$ charges of the particles while
keeping their $U(1)_Y$ charges \cite{kineticmixing}. 
The messengers in this case carry $U(1)^{\prime}$ charges.
It could also modify
the low energy phenomenology. For example, the shifted charges for a specific 
model could be such that
the low energy model is ``leptophobic'', and hence allows a lighter $Z^{\prime}$.
However, the size of the correction
depends in detail on the particle content of the model and on the 
decoupling scales. 
Kinetic mixing is again an ultra-violet effect 
(which, however,  vanishes in the limit $g_{1^{\prime}}/g_{1} \rightarrow 0$) 
and was not considered here in detail.

Lastly, we would like to comment that a straightforward  
extension of our mechanism can lead to lepton number violation thorough couplings
$h_{\not{L}}SLH_{2} \rightarrow \mu_{\not{L}}LH_{2}$ if $L$ and $H_{1}$
carry the same $U(1)$$^{\prime}$ charge (this is the case for the
$U(1)$$_{\eta}$ of $E_6$, but not for the other $U(1)^{\prime}$ embeddings)
or more generally in a multi-singlet model. 
The  case $Q_{H_{1}} \neq Q_{L}$ is in fact more
attractive since it would forbid lepton number violating Yukawa operators 
in the high-energy theory.
Since gauge mediation guarantees slepton-Higgs
mass universality, and Higgs-slepton bilinear mixing in the scalar
potential arises only from radiative  $A$-parameters, then all
conditions for the dynamical alignment suppression of neutrino masses
outlined in Ref.~\cite{NP2} are automatically and naturally 
satisfied and gauge-mediation models
would lead in this case to also a successful generation of neutrino masses. 

In conclusion,
the Higgs mass parameters in the gauge-mediation framework are best
understood as dynamical degrees of freedom corresponding to a (SM) singlet.
Here, it was suggested that such a singlet is not a gauge singlet but
transforms under a $U(1)$$^{\prime}$.
While the minimal (``model-independent'') low energy framework was given and discussed, its many
possible extensions and embeddings remain to be explored in greater
detail.

\acknowledgments

It is pleasure to thank H.~Murayama and especially
M.~Cveti\v c for discussions. 
We also thank J.~Feng and T.~Moroi for their comments on CP violation,
as well as K.~Agashe for his comments.
This work was supported by the U.S. Department of Energy grants
No.~DOE-EY-76-02-3071 and No.~DE-FG02-96ER40559.


\newpage 

\begin{thebibliography}{100}

\bibitem{DNS} 
M.~Dine, A.E.~Nelson and Y.~Shirman,
Phys. Rev. {\bf D51}, 1362 (1995).

\bibitem{FPT} 
J.L.~Feng, N.~Polonsky and S.~Thomas,
Phys. Lett. {\bf B370}, 95 (1996).


\bibitem{DGP} 
G.~Dvali, G.F.~Giudice and A.~Pomarol, 
Nucl. Phys. {\bf B478}, 31 (1996).


\bibitem{NP} 
H.P.~Nilles and N.~Polonsky,
Phys. Lett. {\bf B412}, 69 (1997).


\bibitem{Y} 
T.~Yanagida
Phys. Lett. {\bf B400}, 109 (1997).

\bibitem{CL}
D. Suematsu and Y. Yamagishi, 
Int.~J.~Mod.~Phys. {\bf A10}, 4521 (1995); 
M.~Cvetic and P.~Langacker,
Phys. Rev. {\bf D54}, 3570 (1996);
Mod. Phys. Lett. {\bf A11}, 1247 (1996).


\bibitem{sugra} 
See, for example, 
J.E.~Kim and H.P.~Nilles,
Phys. Lett. {\bf 138B}, 150 (1984);
Mod. Phys. Lett. {\bf A9}, 3575 (1994);
G.F.~Giudice and A.~Masiero,
Phys. Lett. {\bf B206}, 480 (1988);
C.~Kolda, S.~Pokorski and N.~Polonsky,
Phys. Rev. Lett. {\bf 80}, 5263 (1998).

\bibitem{penn} 
G.~Cleaver, M. Cveti\v c, J.R.~Espinosa, L.~Everett, 
P.~Langacker  and J.~Wang, Phys.~Rev. {\bf D59}, 055005 (1999);     
hep-ph/9811355, Phys.~Rev. {\bf D}, to appear. 

\bibitem{P} 
P.~Ciafaloni and A.~Pomarol, 
Phys. Lett. {\bf B404}, 83 (1997).


\bibitem{singlet} 
M.~Dine and A.E.~Nelson, Phys. Rev. {\bf D48}, 1277 (1993);
K.~Agashe and M.~Graesser, Nucl. Phys. {\bf B507}, 3 (1997);
A.~de Gouvea, A.~Friedland and H.~Murayama, Phys. Rev. {\bf D57}, 5676 (1998).

\bibitem{GMRSB} 
See, for example,
S.~Dimopoulos, S.~Thomas and J.D.~Wells,
Nucl. Phys. {\bf B488}, 39 (1997);
J.A.~Bagger, K.~Matchev, D.M.~Pierce and R.~Zhang,
Phys. Rev. {\bf D55}, 3188 (1997).

\bibitem{Moroi}
T.~Moroi, Phys. Lett. {\bf B447}, 75 (1999).

\bibitem{EL} For a recent discussion of the experimental constraints
see J.~Erler and P.~Langacker, hep-ph/9903476.

\bibitem{CDM} 
H.-C.~Cheng, B.A.~Dobrescu and K.T.~Matchev,
Phys. Lett. {\bf B439}, 301 (1998);
Nucl. Phys. {\bf B543}, 47 (1999).


\bibitem{LW}
See, for example,  
P.~Langacker and J.~Wang, Phys. Rev. {\bf D58}, 115010 (1998). 

\bibitem{MRK} 
C.~Kolda and J.~March-Russell,
Phys. Rev. {\bf D55}, 4252 (1997).

\bibitem{kineticmixing} 
See, for example,
K.S. Babu, C. Kolda and J. March-Russell, Phys. Rev. {\bf D54}, 4635 (1996).

\bibitem{NP2}
H.P.~Nilles and N.~Polonsky,
Nucl. Phys. {\bf B484}, 33 (1997).



\end{thebibliography}
\end{document}